# HAC-Net: A Hybrid Attention-Based Convolutional Neural Network for Highly Accurate Protein-Ligand Binding Affinity Prediction

Gregory W. Kyro[†] Rafael I. Brent[†] Victor S. Batista[*]

Department of Chemistry, Yale University

{gregory.kyro, rafi.brent, victor.batista}@yale.edu

[†]authors contributed equally to this work; [*]corresponding author

## Abstract

*Applying deep learning concepts from image detection and graph theory has greatly advanced protein-ligand binding affinity prediction, a challenge with enormous ramifications for both drug discovery and protein engineering. We build upon these advances by designing a novel deep learning architecture consisting of a 3-dimensional convolutional neural network utilizing channel-wise attention and two graph convolutional networks utilizing attention-based aggregation of node features. HAC-Net (**H**ybrid **A**ttention-Based **C**onvolutional Neural **Net**work) obtains state-of-the-art results on the PDBbind v.2016 core set, the most widely recognized benchmark in the field. We extensively assess the generalizability of our model using multiple train-test splits, each of which maximizes differences between either protein structures, protein sequences, or ligand extended-connectivity fingerprints of complexes in the training and test sets. Furthermore, we perform 10-fold cross-validation with a similarity cutoff between SMILES strings of ligands in the training and test sets, and also evaluate the performance of HAC-Net on lower-quality data. We envision that this model can be extended to a broad range of supervised learning problems related to structure-based biomolecular property prediction. All of our software is available as open source at https://github.com/gregory-kyro/HAC-Net/, and the HACNet Python package is available through PyPI.*

# 1. Introduction

## 1.1 Motivation

Protein-ligand interactions are essential for most biomolecular mechanisms, including important processes such as gene regulation, immunoreaction, and signal transduction.[1] Thoroughly understanding such interactions is therefore necessary for many targeted applications such as drug discovery and protein design. Specifically, efficient and accurate screening for particular binding properties would enable quick identification of inhibitory molecules that combat disease and proteins that perform desired functions, and thus a model which can quickly and accurately predict protein-ligand interactions would be incredibly powerful for promoting important molecular-level applications.

The intrinsic complexity of biological data has motivated the use of machine learning (ML) to create models capable of predicting intricate biomolecular phenomena, many of which have proven to be incredibly powerful.[2,3] One notable example is DeepMind's AlphaFold,[4] a revolutionary deep learning model which can accurately predict 3-dimensional protein structure from amino acid sequence. In the case of protein-ligand interactions, ML has enabled the discovery of new drugs and chemicals that would not have seemed intuitive to investigate based on chemical theory alone.[5]

## 1.2 Background

There has been continuous progress in applying ML to predict protein-ligand binding affinity, gaining significant popularity in 2010 with NNScore,[6] an ensemble of 10 multi-layer perceptrons (MLPs), and RF-Score,[7] a random forest-based model. Many groups have subsequently utilized random forest-based approaches[7-11] or related methods such as gradient-boosted trees[12-16] to predict binding affinity, and most other architectures contain one or multiple MLPs as subcomponents. Convolutional neural networks (CNNs) have become increasingly popular for binding affinity prediction due to their success on image detection tasks.[17] CNNs comprise a class of deep learning architectures where the model learns weights for multiple convolutional filters that scan over the input dataset, transforming it into an output feature map. Many CNNs for binding affinity prediction operate on dimensionality-reduced data, and come in the form of either 1- or 2-dimensional CNNs.[18-26] The requirement of lower-dimensional data is removed by the use of 3-dimensional CNNs (3D-CNNs), which utilize a 3-dimensional voxel representation of protein-ligand complexes where each voxel corresponds to an atomic feature vector. Many groups have employed some form of 3D-CNN for binding affinity prediction.[27-33] There have also been successful efforts to predict binding affinity utilizing graph convolutional networks (GCNs).[34-36] In the case of GCNs, protein-ligand complexes are represented as graphs, where nodes usually correspond to atoms and edges are pathways for information transfer between pairs of nodes. Additionally, recent work by Jones et al. has shown that the fusion of a 3D-CNN and a GCN can result in greater performance than either model in isolation.[37]

The use of attention, a context-based weighting technique analogous to cognitive attention, has been shown to improve performance in many deep learning models.[38-43] Squeeze-and-Excitation Networks won the ImageNet Large Scale Visual Recognition Challenge 2017 for image classification, and have been shown to significantly improve accuracy at a minimal increase in computational cost in many high-performing CNNs.[39-41] A squeeze-and-excitation (SE) block incorporates attention by performing channel-wise feature recalibration. The spatial dimensions for each channel are condensed to a single number via average pooling, and then passed through a network of two fully-connected layers with rectified linear unit (ReLU) activation[44] after the first layer and sigmoid activation after the second. Finally, the resulting vector elements are used as multiplicative weights for the corresponding channels of the input data. Gated graph neural networks (GG-NNs) extend upon traditional graph neural networks by incorporating attention to the aggregation of node features.[42] A neural network first computes attention scores used to weight node features, and then the weighted feature sets are summed. This technique has performed exceptionally well on a wide range of problems in graph-based ML.[42,43]

### 1.3 HAC-Net

We build upon these advances by designing a novel deep learning architecture for protein-ligand binding affinity prediction which averages the outputs of a 3D-CNN utilizing channel-wise attention and two GCNs utilizing attention-based aggregation of node features. This combination achieves an optimal balance between the superior performance of our GCNs and the complementary learning style of our 3D-CNN (*SI Appendix*, Fig. S1, Fig. S2). Furthermore, the inclusion of two architecturally identical GCNs mitigates noise resulting from the inherently stochastic nature of the training process. By incorporating multiple forms of attention with advanced concepts from CNN and GCN architectural design, we are able to demonstrate state-of-the-art performance on the PDBbind benchmark for protein-ligand binding affinity prediction, as well as the ability to generalize to complexes unlike those used for training.

## 2. Model Architecture and Theory

### 2.1 Overview

HAC-Net (**H**ybrid **A**ttention-Based **C**onvolutional Neural **Net**work) is a deep learning model composed of one 3D-CNN and two GCNs. The model takes as the inputs oriented protein and ligand structural files and outputs a prediction of the binding affinity between the inputs.

### 2.2 3-Dimensional Convolutional Neural Network

For the 3D-CNN component of HAC-Net, protein and ligand atoms are first embedded into a 3-dimensional spatial grid, each voxel of which corresponds to either a vector of atomic feature elements or 0s, depending on the presence or absence of an atom center, respectively. The input volume dimensions are $48 \times 48 \times 48 \times 19$, where 48 corresponds to the length of each spatial dimension of the voxel grid and 19 corresponds to the number of channels (i.e., the length of the feature vector). This information is presented to the 3D-CNN as a 4-dimensional array. We utilize the atomic feature set first presented by Pafnucy[30]:

- 9 bits (0 or 1) encoding atom types: B, C, N, O, P, S, Se, halogen and metal
- 1 integer (1, 2, or 3) for atom hybridization
- 1 integer counting the number of bonds with heavy atoms
- 1 integer counting the number of bonds with heteroatoms
- 5 bits (0 or 1) encoding hydrophobic, aromatic, acceptor, donor and ring
- 1 float for partial charge
- 1 integer (-1 or 1) denoting either protein or ligand, respectively

While the 3D-CNN makes use of multiple architectural elements (Fig. 1), the most fundamental building block is the convolutional layer.[45] Intuitively, this component creates a linear combination of all channel values in the spatial neighborhood of a given voxel, then propagates the resulting scalar to a corresponding spatial index in the output array (Eq. 1). The coefficients for this linear combination are learned throughout the training and constitute the weights of a filter which is applied uniformly across each of the input voxels. One filter will therefore generate a 3-dimensional output array. By applying multiple independent filters to a given input, the length of the channel dimension of the output can be modulated, where each filter produces a channel of the output.

Each filter is applied over the input signal according to the following equation:

$$\text{output}[x', y', z'] = \text{bias} + \sum_{h=1}^{k_x} \sum_{i=1}^{k_y} \sum_{j=1}^{k_z} \sum_{f=1}^{F} \text{filter}[h, i, j, f] \cdot \text{input}[x + h, y + i, z + j, f] \quad [1]$$

where $k_x$, $k_y$, and $k_z$ are the spatial dimensions of the filter, $F$ is the total number of input channels, and:

$$x = (x' \cdot \text{stride}) + \text{padding} - \lceil k_x/2 \rceil \quad [2]$$

$$y = (y' \cdot \text{stride}) + \text{padding} - \lceil k_y/2 \rceil \quad [3]$$

$$z = (z' \cdot \text{stride}) + \text{padding} - \lceil k_z/2 \rceil \quad [4]$$

We are able to modulate the size of the output feature map by manipulating padding and stride parameters applied to the convolution (according to Eq. 2-4), where padding refers to inserting zeroes around the initial input array, and stride refers to the step size of the filter between each convolution.

The residual layer,[46] which incorporates skip connections between convolutional layers, is an important constituent of the 3D-CNN component of HAC-Net. The outputs of two convolutional layers are summed, and then a subsequent convolutional layer operates on the sum. Architectures containing residual layers are more easily optimized than those relying primarily on standard convolutions, allowing for the training of significantly deeper neural networks which have obtained greatly improved results on standard image-recognition benchmarks.[47-49]

Another key component of the 3D-CNN architecture is the SE block (Fig. 1B), which begins with a standard convolution of the type described above (Eq. 1). The values of each channel are then averaged across all spatial dimensions, yielding a 1-dimensional vector with each index corresponding to a channel:

$$z_c = \frac{1}{L_x \cdot L_y \cdot L_z} \sum_{i=0}^{L_x - 1} \sum_{j=0}^{L_y - 1} \sum_{k=0}^{L_z - 1} \mathbf{u}[i, j, k, c] \quad [5]$$

where $\mathbf{u}$ and $\mathbf{z}$ correspond to the 4-dimensional output of the convolution and the 1-dimensional row vector containing the average value of each channel, respectively. Next, $\mathbf{z}$ is passed through a network of fully-connected layers with ReLU activation after the first layer and sigmoid activation after the second, producing a transformed vector of the same length as the original:

$$\mathbf{s} = \mathrm{sigmoid}(\mathrm{ReLU}(\mathbf{z}\mathbf{W}_1)\mathbf{W}_2) \quad [6]$$

where $\mathbf{W}_1$ $(F \times \frac{F}{16})$ and $\mathbf{W}_2$ $(\frac{F}{16} \times F)$ are the weight matrices for the two fully-connected layers, and $F$ is the number of channels. Finally, each element of $\mathbf{s}$ is used as a multiplicative factor for the corresponding channel of $\mathbf{u}$:

$$\mathbf{x}[i,j,k,c] = s_c\mathbf{u}[i,j,k,c] \quad [7]$$

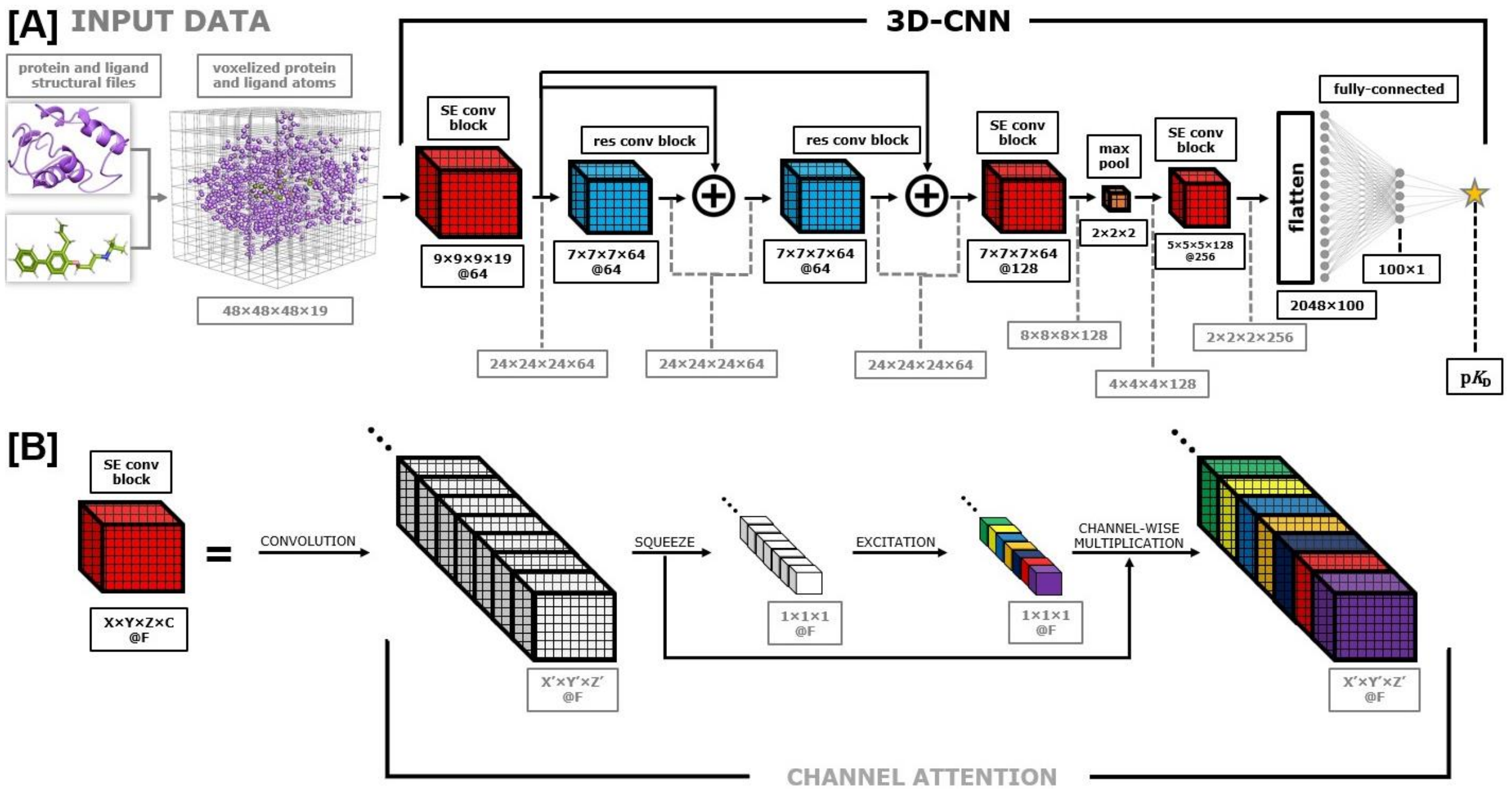

**Figure 1.** 3D-CNN architecture used in HAC-Net. Gray arrows and text refer to data and their transformations. Black arrows and text refer to components of the model architecture. We begin with protein and ligand structural files, voxelize the heavy atoms into a grid of size 48×48×48×19, and then perform a series of convolutions to generate a binding affinity prediction in the form of p$K_D$. The convolutional process is shown in (A). The function of the squeeze-and-excitation (SE) blocks that we incorporate to employ channel-wise attention is visually depicted in (B).

In this way, the model learns to optimally weight the various features based on a transformation of their collective average values, which can be regarded as a self-attention mechanism on the channels.[39]

The complete 3D-CNN training procedure consists of both a feature extraction protocol (Fig. 1A) and subsequent optimization of a fully-connected network (Fig. 5C). The voxelized protein and ligand structural data of size 48×48×48×19 are first passed to an SE block with filter size 9×9×9×19@64, where 64 denotes the number of identical filters, corresponding to the number of channels in the output feature map. It is important to note that all convolutional layers are followed by ReLU activation and batch normalization[50] with momentum of 0.1 for estimating the moving mean and moving variance. The transformed data, now of size 24×24×24×64, are then passed to two residual layers, each of size 7×7×7×64@64. The data are then fed into another SE block of size 7×7×7×64@128, producing an output of size 8×8×8×128. We then apply a max pooling layer which divides the spatial grid into sub-grids of size 2×2×2 for every channel and propagates the maximal value of each one, reducing the size of the data to 4×4×4×128. The data are then passed to a third SE block of size 5×5×5×128@256, downsizing the data to 2×2×2×256. Lastly, the data are flattened into a vector of length 2048 and passed to a fully-connected layer of size 2048×100 with ReLU activation and batch normalization, then to a final fully-connected layer of size 100×1, resulting in a binding affinity output in the form of p$K_D$ (Fig. 1A).

After the initial 3D-CNN training is complete, we extract the flattened features of size 2048 and use them to train a pair of fully-connected layers identical to those in the 3D-CNN architecture other than the single exception of 0.3 momentum used for batch normalization. This protocol notably improves performance, likely due to the fact that the fully-connected layers account for only 1.9% of the total parameters in the 3D-CNN (as compared with 59.5% in the GCNs), causing the initial learning to be driven primarily by the parameters for the convolutional layers. Therefore, independently training fully-connected layers on the extracted features enables them to adapt more precisely to the outputs of the convolutional layers.

## 2.3 Graph Convolutional Networks

The GCN components of HAC-Net interact with the input data in a fundamentally different manner than the 3D-CNN. Rather than using a voxel representation of atoms, the protein-ligand complexes are represented as graphs, where nodes correspond to heavy atoms and edges are pathways for information transfer between the nodes. In the case of HAC-Net, edges exist between atoms whose centers are within 3.5 Å. For the GCNs, we utilize the Pafnucy feature set with the addition of van der Waals radius (float), for a total of 20 atomic features.

GCNs are a broad class of networks which iteratively update node features according to three general steps: message creation, aggregation, and feature updating.[51,52] In HAC-Net, message creation involves a dimensionality-preserving linear transformation applied to each set of node features. For each node, the resulting messages of its neighbors are weighted by the distance from the central node, and then aggregated according to a specified algorithm (Eq. 8). We apply an attention mechanism similar to that of GG-NNs,[42] with the important distinctions that we apply node-wise (rather than channel-wise) attentional weights and we use it in the message-passing aggregation step rather than for generating graph-level features. In our case, the function operates according to the following equation[53,54]:

$$\mathbf{x}_{\mathrm{out}} = \mathrm{softmax}(\mathit{\Gamma}\mathbf{X})^{\mathrm{T}}(\mathit{\Omega}\mathbf{X}) \quad [8]$$

where $\mathbf{X}$ denotes the matrix whose rows are the messages created from the central node and its neighbors. $\mathit{\Gamma}$ and $\mathit{\Omega}$ are independent neural networks, and $\mathit{\Gamma}\mathbf{X}$ indicates application of $\mathit{\Gamma}$ to each row of $\mathbf{X}$ followed by vertical concatenation of the outputs. In HAC-Net, $\mathit{\Gamma}$ is a set of three fully-connected layers (20×10, 10×5, and 5×1) with Softsign activation[55] after the first two, and $\mathit{\Omega}$ is the identity. This operation allows us to weight each node in the message-passing mechanism by a corresponding attentional score.

After message creation and aggregation, the node features are updated by combining the original node features (pre-message creation) with the node features after aggregation. We utilize a simplified gated recurrent unit (GRU)[56] for updating node features, which performs the following operations[57,58]:

$$\text{reset gate: } \mathbf{r} = \text{sigmoid}(\mathbf{W}_{xr}\mathbf{x} + \mathbf{b}_{xr} + \mathbf{W}_{hr}\mathbf{h}_0 + \mathbf{b}_{hr}) \quad [9]$$

$$\text{update gate: } \mathbf{z} = \text{sigmoid}(\mathbf{W}_{xz}\mathbf{x} + \mathbf{b}_{xz} + \mathbf{W}_{hz}\mathbf{h}_0 + \mathbf{b}_{hz}) \quad [10]$$

$$\text{new gate: } \mathbf{n} = \tanh\left(\mathbf{W}_{xn}\mathbf{x} + \mathbf{b}_{xn} + \mathbf{r} \odot (\mathbf{W}_{hn}\mathbf{h}_0 + \mathbf{b}_{hn})\right) \quad [11]$$

$$\text{output gate: } \mathbf{y} = (1 - \mathbf{z}) \odot \mathbf{n} + \mathbf{z} \odot \mathbf{h}_0 \quad [12]$$

where $\mathbf{h}_0$ and $\mathbf{x}$ are the pre- and post-message-passing data, respectively. The matrices $\mathbf{W}_{ij}$ and vectors $\mathbf{b}_{ij}$ denote learnable weights and biases, respectively. $\odot$ indicates element-wise multiplication, and $\mathbf{y}$ denotes the output, which is established as the new vector of node features for the next round of message passing.

Our model performs four iterations of message passing, and then the outputs from the fourth GRU iteration are processed using a method presented by Jones et al.,[37] which we refer to as *asymmetric attentional aggregation*. The operation is performed according to the following equation:

$$\mathbf{y}_{\text{out}} = \sum_{v \in V} \text{softmax}\left(\Gamma(\mathbf{Y}_v || \mathbf{X}_v)\right) \odot (\Omega \mathbf{X}_v) \quad [13]$$

where $\mathbf{Y}_v||\mathbf{X}_v$ denotes horizontal concatenation of the post- and pre-message-passing data, respectively, for node $v$ in the set of all nodes, $V$. $\Gamma$ is a set of two fully-connected layers (40×20 and 20×128), and $\Omega$ is a single linear transformation (20×128), both using Softsign activation. The output of asymmetric attentional aggregation ($\mathbf{y}_{\text{out}}$) is then passed through a final set of three fully-connected layers (128×85, 85×64, and 64×1), the first two of which are followed by ReLU activation, to generate a binding affinity prediction. The GCN process is visually depicted in Figure 2.

## 3. Data

The PDBbind database[59] is an online repository of experimentally determined binding affinity data for biomolecular complexes deposited in the Protein Data Bank. In this work, we make use of the protein-ligand complexes contained in the PDBbind v.2020 database (19,443 total complexes). For each protein-ligand complex, the protein and protein pocket are provided in PDB format, where the protein pocket is defined as all of the amino acid residues within 10 Å of the ligand. Ligand coordinates are provided in both MOL2 and SDF formats, and the associated binding affinity is given as either $pK_D$ or $pK_I$. In this work, we utilize only the MOL2 ligand files. In most cases, the 3-dimensional structures of protein-ligand complexes are determined by crystallography, although there are also relatively few cases where structures are determined with nuclear magnetic resonance (NMR) spectroscopy (Fig. 3C). The PDBbind v.2020 refined set contains 5,316 data points with high-quality labels and structures, as identified by the PDBbind team according to a rigorous set of requirements.[60]

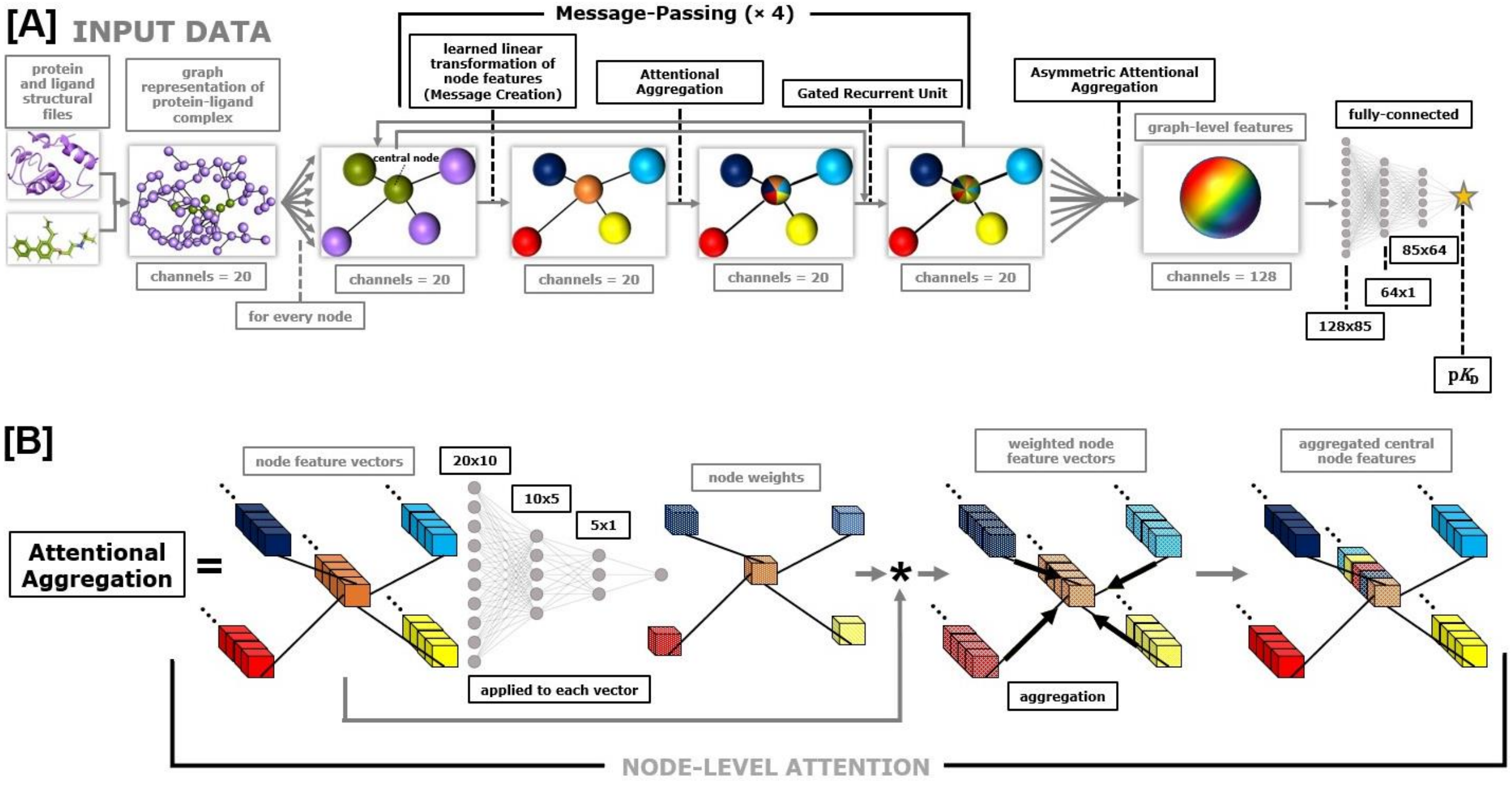

**Figure 2.** GCN architecture used in HAC-Net. Gray arrows and text refer to data and their transformations. Black arrows and text refer to components of the model architecture. We begin with protein and ligand structural files, represent the heavy atoms as a graph, and then perform four iterations of message passing, after which we employ asymmetric attentional aggregation to generate a graph-level feature vector. Finally, the data are passed through a series of fully-connected layers to yield a prediction of the protein-ligand binding affinity in the form of p$K_D$. The full GCN protocol is shown in (A). The function of attentional aggregation that we incorporate to employ node-level attention is visually depicted in (B).

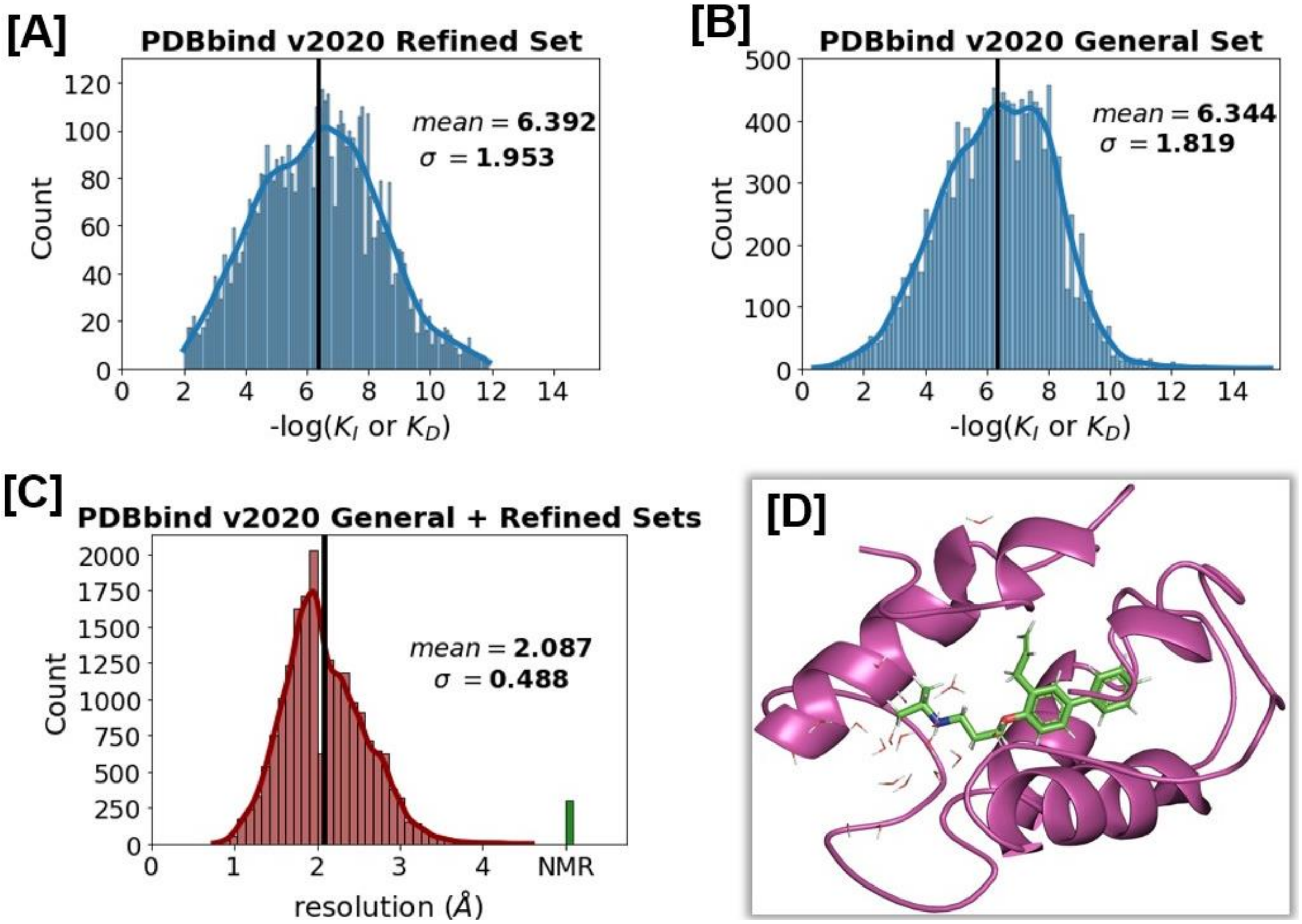

**Figure 3.** Characteristics of the PDBbind v.2020 protein-ligand database. (A) Distribution of binding affinity labels for the refined set. (B) Distribution of binding affinity labels for the general set (excluding the refined set). (C) Distribution of crystal structure resolution (red) and nuclear magnetic resonance (NMR) data points (green) for the general set (entire PDBbind v.2020 protein-ligand database). (D) Representative protein ligand complex (PDB ID: 3ACX), protein is shown as magenta, ligand is shown as light green.

Moreover, we utilize the PDBbind v.2016 core set as a test set, which is a collection of 290 complexes chosen from a wide distribution of structural clusters and binding affinities. This benchmark, inspired by the 2016 Comparative Assessment of Scoring Functions (CASF-2016) test set,[61] is the most widely reported benchmark for protein-ligand binding affinity prediction.

## 4. Performance on the PDBbind v.2016 Core Set Crystal Structures

### 4.1 Comparison to Existing Models

We test and report results on the PDBbind v.2016 core set in Table 1 to directly compare HAC-Net to the highest-performing models in the literature (to the best of our knowledge). It is important to note that for all HAC-Net results presented in this work, there is no overlap between training, validation and test sets, and model hyperparameters were optimized exclusively on training and validation data. HAC-Net achieves the lowest root-mean-square error (RMSE) among models reported in the literature (Fig. 4), as well as the highest Spearman $\rho$ and $r^2$ values. Furthermore, our model attains the second-highest

**Table 1.** Comparison to high-performing models for predicting protein-ligand binding affinity on crystal structures of the PDBbind v.2016 core set.

| **Model** | **RMSE** | **MAE** | **$r^2$** | **Pearson r** | **Spearman $\rho$** |
|---|---|---|---|---|---|
| ***HAC-Net*** | **1.205** | 0.971 | **0.692** | 0.846 | **0.843** |
| TopBP* [62] | 1.210 | N/R | N/R | **0.861** | N/R |
| AEScore[63] | 1.22 | N/R | N/R | 0.83 | 0.64 |
| AK-score* [31] | 1.22 | N/R | N/R | 0.812 | 0.670 |
| DeepAtom[29] | 1.232 | **0.904** | N/R | 0.831 | N/R |
| *HAC-Net*[‡] | 1.259 | 1.020 | 0.664 | 0.819 | 0.814 |
| PerSpect ML* [14] | 1.265 | N/R | N/R | 0.840 | N/R |
| $K_{DEEP}$* [28] | 1.27 | N/R | N/R | 0.82 | 0.82 |
| AGL-Score* [13] | 1.272 | N/R | N/R | 0.833 | N/R |
| OnionNet[20] | 1.278 | 0.984 | N/R | 0.816 | N/R |
| PSH-GBT* [16] | 1.280 | N/R | N/R | 0.835 | N/R |
| FAST[37] | 1.308 | 1.019 | 0.638 | 0.810 | 0.807 |
| BAPA[23] | 1.308 | 1.021 | N/R | 0.819 | 0.819 |
| SIGN[35] | 1.316 | 1.027 | N/R | 0.797 | N/R |
| TopologyNet* [18] | 1.34 | N/R | N/R | 0.81 | N/R |
| DockingApp RF* [11] | 1.35 | 1.09 | N/R | 0.83 | N/R |
| DeepDTAF[24] | 1.355 | 1.073 | N/R | 0.789 | N/R |
| DLSSAffinity[33] | 1.40 | N/R | N/R | 0.79 | N/R |
| Pafnucy[30] | 1.42 | 1.13 | N/R | 0.78 | N/R |
| Pair[64] | 1.44 | N/R | N/R | 0.75 | N/R |
| GraphBAR[36] | 1.542 | 1.241 | N/R | 0.726 | N/R |
| PointTransformer[12] | 1.58 | 1.29 | N/R | 0.753 | 0.751 |
| MGNN* [65] | N/R | N/R | N/R | 0.85 | N/R |
| SE-OnionNet[22] | N/R | N/R | N/R | 0.83 | N/R |
| PLEC-NN* [10] | N/R | N/R | N/R | 0.817 | N/R |

[a] Root-mean-square error (RMSE) in units of p$K_D$, mean absolute error (MAE) in units of p$K_D$, $r^2$, Pearson r, and Spearman $\rho$ are shown.
[b] Models are ranked by RMSE in increasing order.
[c] The best value for each metric is shown in bold.
[d] * indicates that the model did not use a validation set, which is expected to present overly optimistic results.
[e] HAC-Net is trained on the PDBbind v.2020 general set; HAC-Net‡ is trained on the PDBbind v.2016 refined set.
[f] Models trained on PDBbind v.2016 refined set: TopBP, AEScore, AK-score, PerSpectML, $K_{DEEP}$, AGL-Score, PSH-GBT, BAPA, SIGN, TopologyNet, GraphBAR, PointTransformer, MGNN; Models trained on PDBbind v.2016 general set: DeepAtom, OnionNet, FAST, DeepDTAF, DLSSAffinity, Pafnucy, PLEC-NN; Models trained on PDBbind v.2018 refined set: Pair; Models trained on PDBbind v.2018 general set: DockingApp RF, SE-OnionNet.

Pearson r and the second-lowest mean absolute error (MAE) in the field.

It is important to note that for the results presented in Table 1, the highest performing version of HAC-Net is trained on all complexes in the PDBbind v.2020 general set that do not appear in either the v.2016 core set or our randomly generated validation set. The other models presented have been trained on a variety of different training sets, including the PDBbind v.2016 refined set, the v.2016 general set, the v.2018 refined set, and the v.2018 general set. Given that the performance of deep learning models is often improved when trained on more data, it is important to note that models trained with less data may indeed perform better on the PDBbind v.2016 core set benchmark if they were trained on more data. However, it is also the case that for some models, training on only the refined set actually improves performance on the core set benchmark,[37] given that the protein-ligand complex data in the refined set are of higher quality than those in the general set and the core set complexes are selected from the refined set.[60] This caveat, along with many others that we elucidate in later sections, necessitates training and evaluating on multiple train-test splits, as we demonstrate rigorously in later sections.

Additionally, we utilize the procedures from CASF-2016[61] to assess the ability of HAC-Net to accurately rank, dock, and screen for ideal protein-ligand pairs. We report our results, along with the complete procedures followed, as Supporting Information (*SI Appendix*, Table S3). Our model does not achieve comparable performance to commonly-used docking programs such as AutoDock Vina,[66] which is unsurprising given that HAC-Net is not explicitly optimized for such tasks. However, many docking and screening approaches suffer from lower Pearson r and higher RMSE values compared to high-performing ML-based approaches (Table 1).[61] Several groups have integrated these traditional docking methods with modern ML approaches,[32,63,67] and it has been shown that such combinations can largely retain both the precision of the ML-based component and the docking/screening power of the classical component, strongly motivating the parallel optimization of both components.[62]

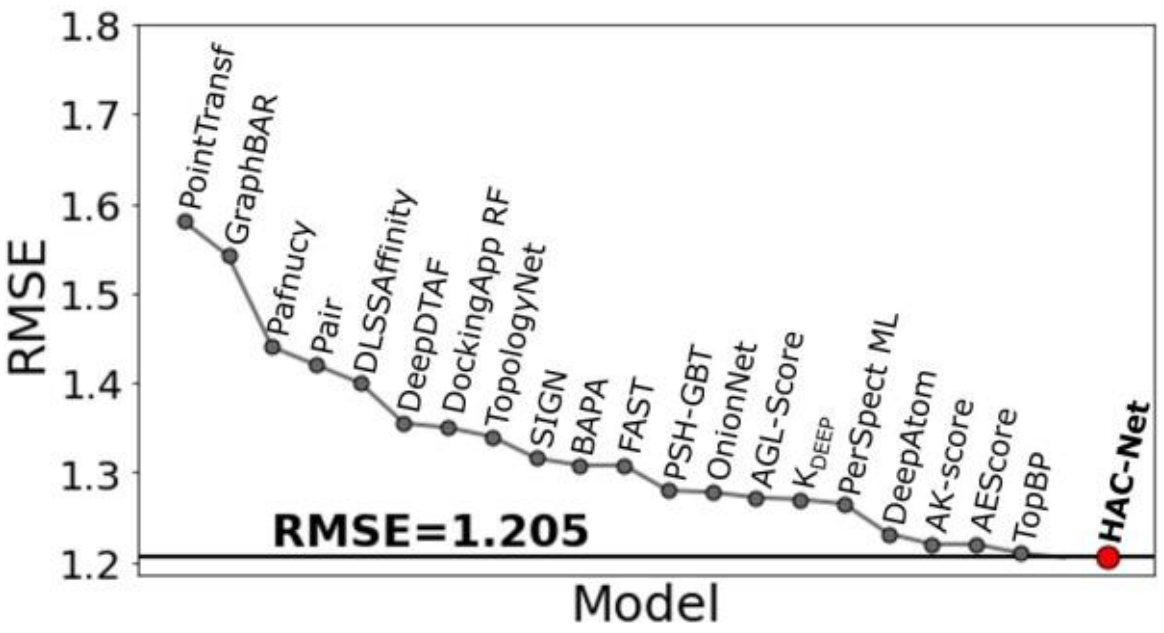

**Figure 4.** Comparison of root-mean-square error (RMSE) provided in units of p$K_D$ on the PDBbind v.2016 core set benchmark across all high-performing models in the literature (to the best of our knowledge) for protein-ligand binding affinity prediction. HAC-Net achieves the lowest RMSE with a value of 1.205 p$K_D$.

## 4.2 Improved Performance Due to Attention-Based Implementations

To demonstrate the importance of our model's attention-based components, we independently train and test an analogous model without SE blocks in the 3D-CNN and node-wise attentional aggregation in the GCNs, which we refer to as *vanilla HAC-Net.* We use identical training and validation sets as those used for HAC-Net to provide an accurate comparison of performance. Results on the PDBbind v.2016 core set are shown in Figure 5A-B, and it is clear that the inclusion of attention-based components significantly improves the performance of our model.

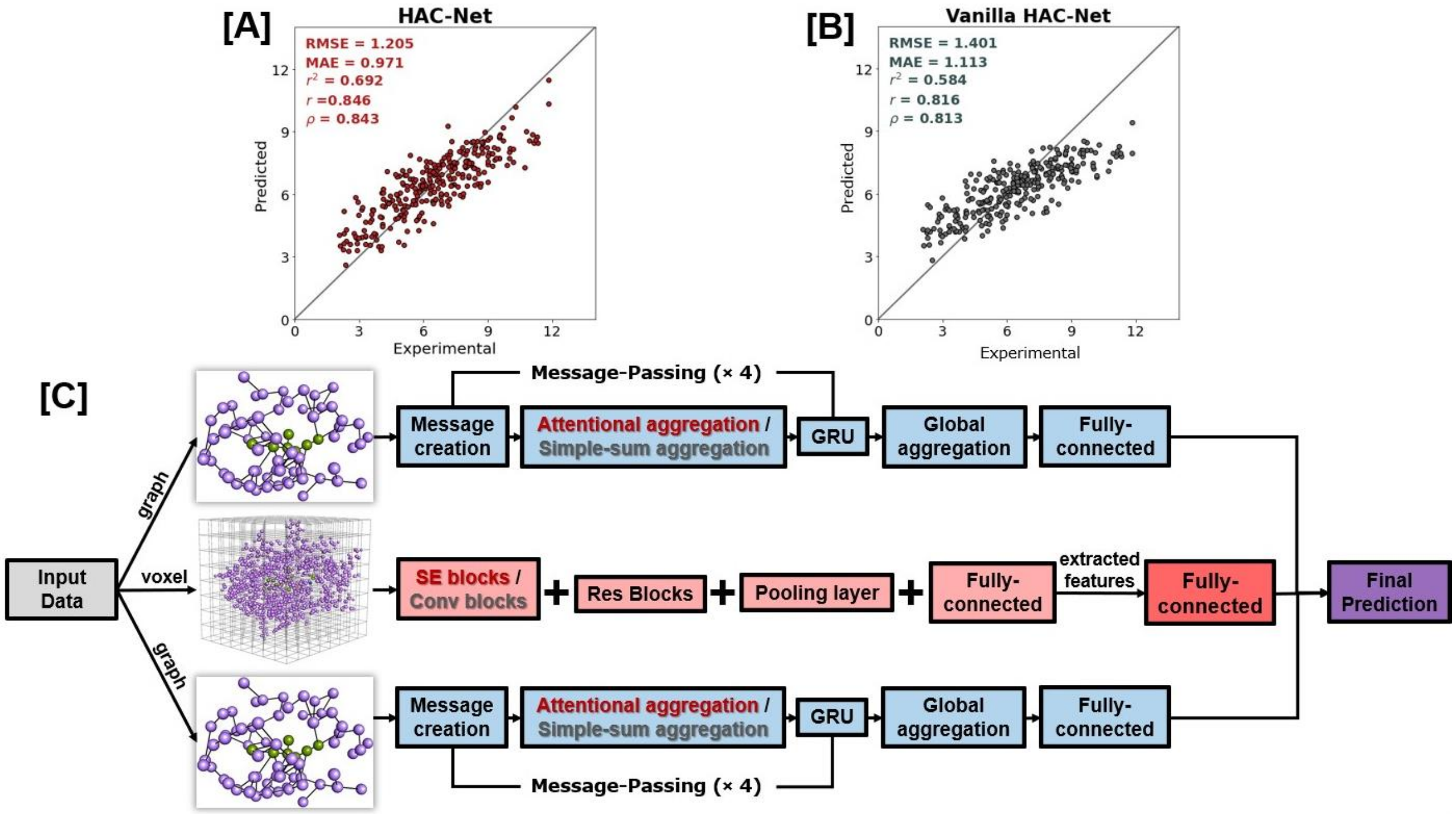

**Figure 5.** HAC-Net performance on PDBbind v.2016 core set benchmark and basic architectural scheme. (A) Correlation scatter plot depicting HAC-Net predictions of experimental p$K_D$ values for core set complexes. (B) Correlation scatter plot depicting vanilla HAC-Net predictions of experimental p$K_D$ values for core set complexes. Root-mean-square error (RMSE) provided in units of p$K_D$, mean absolute error (MAE) in units of p$K_D$, $r^2$, Pearson r, and Spearman $\rho$ are shown for A and B. (C) Basic scheme of HAC-Net architecture. Blue blocks denote components of the GCNs, red blocks denote components of the 3D-CNN. In the GCNs, HAC-Net utilizes attentional aggregation (red), while vanilla HAC-Net utilizes simple-sum aggregation (gray). In the 3D-CNN, HAC-Net utilizes squeeze-and-excitation (SE) blocks (red), while vanilla HAC-Net utilizes ordinary convolutional blocks (gray).

## 5. Demonstration of Generalizability

### 5.1 Concerns With Deep Learning Models for Predicting Protein-Ligand Binding Affinity

There is significant concern in the literature regarding the inability of many deep learning models for binding affinity prediction to successfully generalize to data that are dissimilar to what they have been trained on.[68] Moreover, the PDBbind database, which is the most commonly used database for protein-ligand binding affinity prediction, contains an appreciable bias due to the preferential tendency of experimentalists to measure certain classes of complexes that have been deemed worthy of investigation. Although this pattern in the data creates difficulty in training a model which can comprehensively sample chemical space, the data can be used to train models aimed at predicting binding affinities for the most interesting complexes (i.e., those with suspected biological and/or pharmaceutical significance, etc.).

Volkov et al. have suggested that most models for predicting binding affinity learn primarily via memorization rather than by modeling any physically meaningful phenomena.[68] To demonstrate this effect with a standard graph neural network, they trained and tested using only the proteins or only the ligands, and achieved results comparable to those obtained when the model was trained with the protein-ligand complexes. Moreover, this effect has been previously observed by others.[69] We therefore train HAC-Net with protein-only and ligand-only data, and test performance on the PDBbind v.2016 core set to elucidate the extent to which this effect is present in our model. We find that performance on the core set is significantly worse in the cases of protein-only and ligand-only data compared to using complete protein-ligand complexes (Table 1), although there is clearly a degree of memorization involved as our model achieves nontrivial results (*SI Appendix*, Fig. S4). For training and testing on protein-only data, RMSE increases drastically from 1.205 to 1.742 and Pearson r decreases from 0.846 to 0.638. In the case of ligand-only data, RMSE increases to 1.605 and Pearson r decreases to 0.741.

Multiple groups have suggested that the high performance of many deep learning models for predicting protein-ligand binding affinity is predicated on the similarity of proteins and ligands in the training and test sets used.[68-70] For example, Li et al. have shown that imposing template modeling (TM)-score and sequence identity cutoffs between proteins in the training and test sets can lead to significant reductions in performance of deep learning models for binding affinity prediction.[70] Therefore, it is incumbent on new work in this field to demonstrate generalizability to complexes unlike those used for training, such that there is legitimate applicability to studying new systems. In this work, we present a rigorous protocol for doing exactly this. We illustrate the ability of our model to generalize across protein structure by utilizing agglomerative hierarchical clustering[71] to group proteins by pairwise TM-scores[70] and create training and test sets that maximize differences in protein structures. We perform an analogous evaluation for protein sequence, clustering proteins based on pairwise sequence identity of Needleman-Wunsch (NW)-aligned[70] sequences. In order to account for ligand similarity, we utilize the Butina clustering method[72] based on extended-connectivity fingerprints[73] with Tanimoto coefficient ($T_c$) cutoffs[74] to maximize the differences between ligands in the training and test sets. Furthermore, we perform 10-fold cross-validation with a $T_c$ cutoff between ligand SMILES[75] strings in the training and test sets and compare to a control. Lastly, to demonstrate that our model is not specialized for high-quality data like those in the PDBbind v.2016 core set, we evaluate the performance of HAC-Net on lower-quality data from the PDBbind v.2020 general set.

## 5.2 Generalizability Across Proteins Based on Structure and Sequence Similarities

To demonstrate the generalizability of our model to dissimilar proteins, we utilize pairwise structural and sequence homology of the proteins as distance metrics for hierarchical agglomerative clustering. Training, validation, and test sets are then generated from different clusters of the data, ensuring maximal dissimilarity between the proteins of different sets.

Structural similarity between two proteins is defined by the TM-score[70] according to the following equation:

$$\mathrm{TM} = \max\left[\frac{1}{L_t}\sum_{i}^{L_a}\frac{1}{1+\left(\frac{d_i}{d_0(L_t)}\right)^2}\right] \quad [14]$$

where $L_t$ is the length of the test protein, $L_a$ is the number of aligned residue pairs identified by TM-align,[76] $d_i$ is the distance between the $i$th pair of α-carbon atoms of the two structures, and $d_0(L_t) = 1.24\sqrt[3]{L_t - 15} - 1.8$ (a scale that normalizes distances). TM-score is therefore in the range [0,1], where higher values indicate greater similarity between protein structures. In the case of multi-chain proteins, these comparisons are carried out pairwise between all inter-protein chain combinations and the lowest similarity value is recorded.

Protein sequence similarity is determined by aligning the two sequences using the NW algorithm[70] and then computing the sequence identity (i.e., the number of aligned identical residues divided by the length of the longer protein).

In both cases, agglomerative hierarchical clustering is used to create dissimilar groups of protein-ligand complexes based on either protein structure or protein sequence similarity. This unsupervised learning method is initiated with a set of pairwise distances between data points (quantified by either TM-score or sequence identity) and iteratively merges the two most similar clusters.[71] When two existing clusters are merged, the new inter-cluster distances are calculated according to Ward's minimum variance objective function:

$$d(u,v) = \sqrt{\frac{|v|+|s|}{T}d(v,s)^2 + \frac{|v|+|t|}{T}d(v,t)^2 - \frac{|v|}{T}d(s,t)^2} \quad [15]$$

where the clusters $s$ and $t$ have been merged to create a new cluster $u$, and the new distance between $u$ and some cluster $v$ needs to be determined. $|c|$ defines the number of data points in cluster $c$, and $T = |s| + |t| + |v|$.

To promote standardization in the field, we utilize train-test splits provided by Feinberg et al.[34] that were generated with complexes in the PDBbind v.2007 refined set according to the method detailed above, excluding a few complexes which were subsequently removed from the PDBbind database due to quality-control concerns. For structure-based clustering, the training, validation, and test sets contain 919, 256, and 117 complexes, respectively. For sequence-based clustering, the training, validation, and test sets contain 971, 220, and 101 complexes, respectively. As a control, we test on the PDBbind v.2007 core set (209 complexes), validate on 200 complexes from the

PDBbind v.2007 refined set, and train on the remaining 883 complexes from the refined set. The results are presented in Table 2, and it is clear that the protein clustering techniques do not significantly impair the performance of HAC-Net, supporting its ability to generalize to unseen data with respect to protein structure and sequence (*SI Appendix*, Fig. S5).

**Table 2**. Performance on the PDBbind v.2007 core set (Control), test set based on protein structure similarity (Structure-based), and test set based on protein sequence similarity (Sequence-based).

| Test set | RMSE | MAE | $r^2$ | Pearson r | Spearman $\rho$ |
|---|---|---|---|---|---|
| **Control** | 1.447 | 1.153 | 0.598 | 0.807 | 0.824 |
| **Structure-based** | 1.472 | 1.190 | 0.608 | 0.799 | 0.800 |
| **Sequence-based** | 1.301 | 0.980 | 0.583 | 0.796 | 0.775 |

[a]Root-mean-square error (RMSE) in units of p$K_D$, mean absolute error (MAE) in units of p$K_D$, $r^2$, Pearson r, and Spearman $\rho$ are shown.

## 5.3 Generalizability Across Ligands Based on Extended-Connectivity Fingerprint Similarity

In order to assess the generalizability of our model to dissimilar ligands, we cluster ligand extended-connectivity fingerprints up to four bonds (ECFP4s) according to the Butina unsupervised clustering algorithm[69,72] with a $T_c$ cutoff of 0.8. In this case, any two ligands whose pairwise $T_c$ is greater than or equal to 0.8 are considered to be neighbors. The ligands are then ranked by total number of neighbors in descending order, and the first ligand is clustered with all of its neighbors. All ligands within this cluster are then deleted from the remaining list, and cannot serve as either cluster centroids or members of another cluster. A new cluster is then created analogously from the highest-ranked ligand remaining in the list, and the process is iterated until no ligands remain. The smallest of the resulting clusters are combined to make a test set (1182 complexes), the next smallest clusters are assembled into the validation set (1181 complexes), and all remaining complexes are used as the training data (9448 complexes). This protocol ensures that the ligands in the test set are internally diverse and maximally dissimilar to those in the training set. To promote standardization in the field, we utilize the clusters obtained from the PDBbind v.2015 database by Yang et al.[69] using this protocol, and remove the complexes that were discarded by PDBbind due to quality-control concerns.

This protocol yields the following results: RMSE of 1.240, MAE of 0.978, $r^2$ of 0.355, Pearson r of 0.597, and Spearman $\rho$ of 0.527 (*SI Appendix* Fig. S6). It is clear that while the correlation values are considerably reduced by this generalization method, the error values are not significantly impacted. This discrepancy may be explained by the model's use of mean squared error (MSE) as the loss function, which explicitly prioritizes the minimization of error in the training process rather than maximizing correlation.

We see that the ligand-based clustering method used in this work for creating training, validation and test sets evidently hinders the model's performance more significantly than the protein-based methods (Table 2). This occurrence may be partially explained by the model's ability to more effectively learn trends among the ligands than among the proteins, as supported by the greater performance of the model when trained and tested on only ligands as opposed to when only the proteins were used. The relatively low increases in RMSE and MAE metrics show unambiguously that the model is successful to a significant extent, suggesting that the high performance of HAC-Net cannot be attributed to high ligand similarity between training, validation and test sets, and supporting its ability to generalize to unseen data with respect to ligand ECFP4s.

## 5.4 10-Fold Cross-Validation Based on Ligand SMILES Dissimilarity

To further demonstrate the generalizability of HAC-Net, we perform 10-fold cross-validation. Specifically, we generate ten non-overlapping 500-complex test sets from the PDBbind v.2020 refined set, and for the purpose of generating corresponding training and validation sets, we discard any remaining complexes with ligands that do not satisfy certain dissimilarity requirements relative to each test set. As a metric for ligand similarity, we compute the $T_c$ between the SMILES strings of each ligand pair in the PDBbind v.2020 refined set, asserting that no ligands in either the training or validation sets have a $T_c$ greater than or equal to 0.7 with any ligand in the test set. Additionally, we ensure that no ligands in either the training or validation sets have an average $T_c$ greater than 0.25 with all of the ligands in the test set. All validation sets contain 200 complexes, while the training sets have sizes in the range of 2804 to 2945 complexes, with the size variability due to the different ligand identities in the various test sets. An additional evaluation is performed on the PDBbind v.2016 core set to serve as a control, using training and validation sets containing 3100 and 200 complexes, respectively, derived from the PDBbind v.2020 refined set. As can be seen in Table 3, the cross-validation control results are inferior to those presented in Table 1, likely due to the six-fold reduction in the size of the training set.

**Table 3**. Results of 10-fold cross-validation with complexes from the PDBbind v.2020 refined set, asserting that no ligands in either the training or validation sets have a $T_c$ greater than or equal to 0.7 with any ligand in the test set, and that no ligands in either the training or validation sets have an average $T_c$ greater than 0.25 with all of the ligands in the test set.

| Test set | RMSE | MAE | $r^2$ | Pearson r | Spearman $\rho$ |
|---|---|---|---|---|---|
| **Control** | 1.432 | 1.132 | 0.567 | 0.761 | 0.766 |
| **Mean (σ)** | 1.473 (0.028) | 1.170 (0.024) | 0.430 (0.022) | 0.665 (0.013) | 0.670 (0.013) |
| **Set 1** | 1.462 | 1.152 | 0.443 | 0.672 | 0.678 |
| **Set 2** | 1.450 | 1.176 | 0.448 | 0.676 | 0.672 |
| **Set 3** | 1.451 | 1.159 | 0.447 | 0.669 | 0.675 |
| **Set 4** | 1.497 | 1.163 | 0.414 | 0.659 | 0.661 |
| **Set 5** | 1.481 | 1.182 | 0.424 | 0.654 | 0.656 |
| **Set 6** | 1.454 | 1.158 | 0.444 | 0.678 | 0.677 |
| **Set 7** | 1.501 | 1.187 | 0.409 | 0.650 | 0.658 |
| **Set 8** | 1.483 | 1.182 | 0.424 | 0.661 | 0.680 |
| **Set 9** | 1.527 | 1.219 | 0.383 | 0.639 | 0.649 |
| **Set 10** | 1.426 | 1.123 | 0.465 | 0.686 | 0.697 |

[a]Root-mean-square error (RMSE) provided in units of $pK_D$, mean absolute error (MAE) in units of $pK_D$, $r^2$, Pearson r, and Spearman $\rho$ are shown. The average metrics across the ten cross-validation trials are presented as Mean (σ), where σ is the standard deviation.

We find that the error-based metrics (RMSE and MAE) are minimally affected on the cross-validation splits compared to the control, despite nontrivial reductions in correlation-based statistics (Pearson r and Spearman $\rho$). These results largely reinforce the conclusions derived from evaluation on the ECFP4-based test set, namely that minimizing ligand similarity between training and test sets reduces performance but that a significant amount of learning takes place (Table 3). Additionally, the extraordinarily low standard deviations (average coefficient of variation between all five metrics is 0.026) clearly demonstrate the reproducibility of HAC-Net trainings, suggesting that a retrained version of HAC-Net can reliably be expected to meet the same standard of performance.

## 5.5 Performance on Lower-Quality Data Points

The PDBbind v.2016 core set compiles crystal structures of high quality and with high-confidence binding affinity labels.[59,60] However, to determine how the model performs when given lower-quality data for testing, we evaluate HAC-Net with an additional train-test split containing complexes selected from the PDBbind v.2020 general set. The training set contains 18,108 complexes, the validation set contains 300 complexes, and the test set contains 1000 complexes. In particular, 73.67% of the validation set complexes and 75.30% of the test set complexes are not in the refined set, ensuring that the quality of crystal structures used in these sets more precisely reflects the composition of the PDBbind database as a whole (72.62% of total complexes are not in the refined set). The results of this trial are shown in Figure 6, demonstrating the ability of HAC-Net to make accurate predictions for data points of lower quality. Similarly to the performance on both the ligand ECFP4-based test set and the 10-fold cross-validation test sets, the significant drop in correlation metrics but trivial increase in error metrics may be attributed to the use of MSE as the model's loss function. To analyze the results in more detail, we also calculate absolute error for each data point in the test set, and plot these values as a function of crystal structure resolution (Fig. 6B).

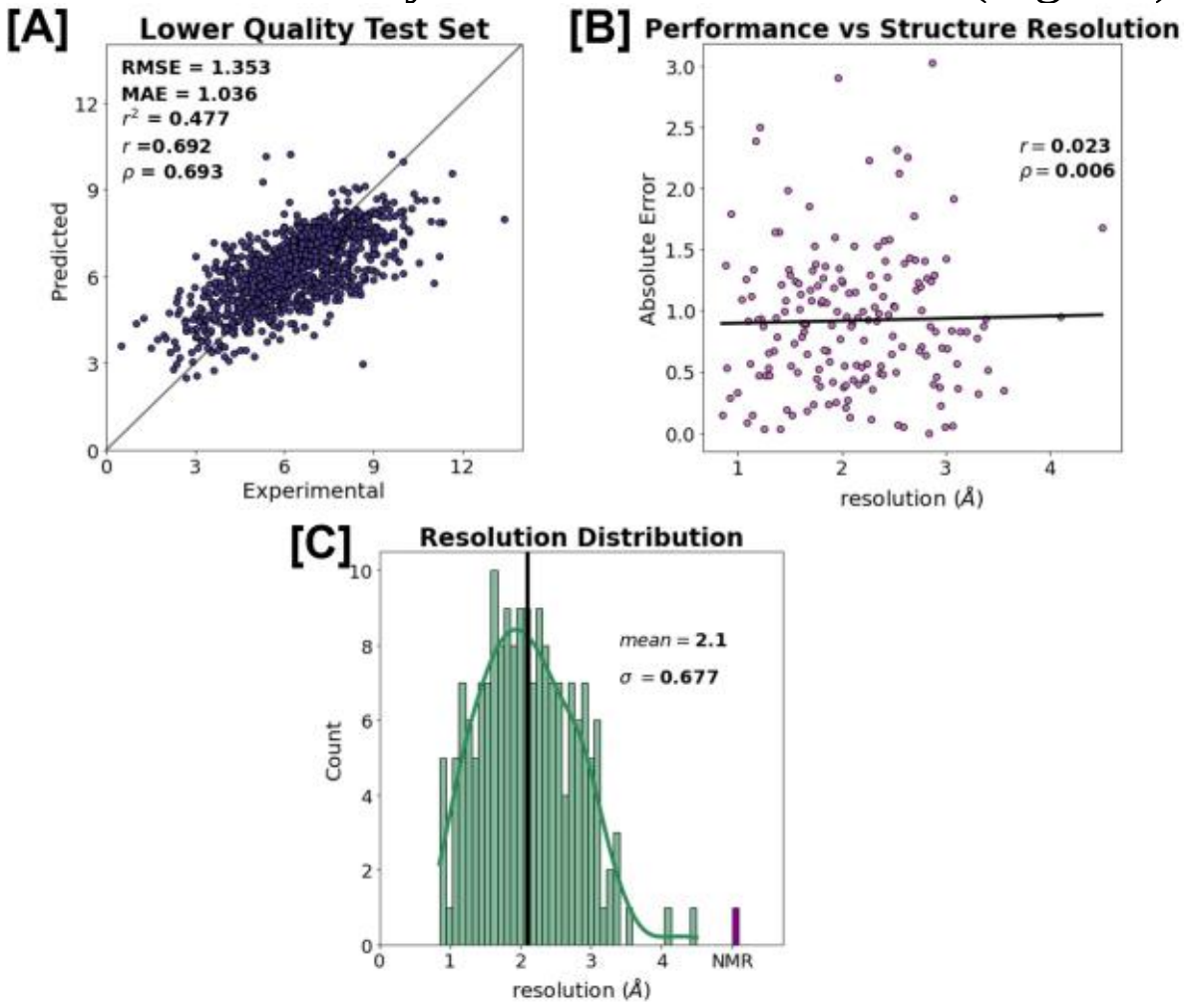

**Figure 6**. HAC-Net performance on lower-quality test set obtained from the PDBbind v.2020 general set. (A) Correlation scatter plot depicting HAC-Net predictions of experimental p$K_D$ values for structures in the lower quality test set. (B) Correlation scatter plot of absolute error for each data point in the test set as a function of crystal structure resolution. Pearson r and Spearman $\rho$ are shown. (C) Histogram showing the distribution of protein-ligand complexes in the lower-quality test set as a function of crystal structure resolution (green) as well as the number of structures determined with NMR (magenta). The mean and standard deviation ($\sigma$) for crystal structure resolution are shown.

The negligible correlation in Figure 6B suggests that there is little or no relationship between structural resolution and the performance of HAC-Net, indicating that its usefulness extends beyond the 2.5 Å resolution required for admission into the refined set. The drop in performance observed when testing on complexes that are excluded from the refined set is thus likely attributable to the other entry requirements for the refined set, which ensure that only the highest-quality structures and binding affinities are included. It is natural that the model would struggle to perform well on lower-quality data, and researchers making use of HAC-Net should consult the criteria for admission into the refined set (other than structural resolution) for details regarding these potential deficiencies.[60] Importantly, these results indicate that HAC-Net is not specialized for high-quality structures like those in the PDBbind v.2016 core set benchmark, and is instead generalizable to crystal structures across a wide range of resolutions and to NMR structures without significantly compromising performance.

## 6. Summary

We have developed HAC-Net, a deep learning model for highly accurate protein-ligand binding affinity prediction. By incorporating multiple forms of attention into our model's architecture, specifically SE blocks into the 3D-CNN and attentional aggregation of node features into the GCNs, we obtain a significant increase in performance. HAC-Net obtains state-of-the-art results on the PDBbind v.2016 core set, the most widely recognized benchmark in the field. We evaluate the generalizability of our model using multiple train-test splits, each of which maximizes differences between either protein structures, protein sequences, or ligand extended-connectivity fingerprints of complexes in training and test sets. Additionally, we perform 10-fold cross-validation with a similarity cutoff between SMILES strings of ligands in the test sets and the corresponding training and validation sets, and also evaluate the performance of HAC-Net on lower-quality data. We demonstrate that our model can successfully generalize to protein-ligand complexes dissimilar to those in the training set, and is not specialized for only high-quality structures.

## 7. Methods

### 7.1 Data Preprocessing

All of the data that we supply to the model were initially downloaded from the PDBbind website.[59] Specifically, we downloaded the PDBbind v.2020 general-except-refined-set and refined-set, both of which contain aligned protein (PDB format), protein pocket (PDB format), and ligand (MOL2 and SDF format) structural files, as well as the corresponding binding affinity data (as either p$K_D$ or p$K_I$). We used the Chimera 1.16 software package[77] to add hydrogens to each protein pocket PDB file and then convert each to MOL2 format. Next, we reformatted any atoms in TIP3P format to avoid compatibility issues with the Pybel software package,[78] which we later used for featurization. Four complexes (PDB IDs: 1A09, 4GII, 4BPS, 4MDQ) could not be interpreted through this process and were therefore discarded. Finally, we used Atomic Charge Calculator II (ACC2)[79] to calculate and add partial charges to each protein pocket MOL2 file. ACC2 determines which method and parameters are suitable for each input structure. The atomic charges for all but five protein pockets were calculated using the Extended Charge Equilibration Method (EQeq).[80] The charges for the protein pockets of PDB IDs 4JDA, 5X5G, 6B8Y, and 4Y16 were calculated using the Charge Equilibration (QEq) method with the parameters presented by Rappé and Goddard.[81] The atomic charges for the protein pocket of PDB ID 5U2F were calculated using the Electronegativity Equalization Method[82] with Raček 2016 (ccd2016_npa) parameters.[83] It is important to note that ACC2 will rarely supply charge estimates which are unreasonable. To account for this, we removed the complexes containing at least one atom with a partial charge assignment of magnitude greater than 2.0 (34 of 19,438 complexes; 0.17%).

## 7.2 Featurization

We processed all of the MOL2 files with the Featurizer module from the tfbio software package[84] to compute features for all heavy atoms. This package utilizes Pybel[78] to collect each heavy atom's atomic number (one-hot encoding), number of bonds with heavy atoms (integer), number of bonds with heteroatoms (integer), hybridization state (integer), and partial charge (float). Next, each molecule's SMARTS[85] string was used to determine whether each atom is hydrophobic, aromatic, a hydrogen bond acceptor, a hydrogen bond donor, and/or a ring (one-hot encoding for each). Additionally, we assigned each atom as present in either the protein or the ligand (-1 or 1, respectively). The features of each atom were then appended to its 3-dimensional coordinates, and the resulting vectors were concatenated vertically to create matrices representing each complex. All of these arrays were then assembled into a file in HDF format and tagged with the corresponding PDB ID and binding affinity label. For use in the GCN components of HAC-Net, an ordered list of van der Waals radii (float) was appended to the HDF file.

## 7.3 Creation of Training, Validation, and Test Splits

After the initial HDF file was generated, the contained data were then partitioned into the training, validation, and test sets used throughout this work. All sets that were selected from previous work (PDBbind v.2016 core set,[61] splits obtained through protein structure- and sequence-based clustering[34], as well as splits obtained through ECFP4-based clustering[69]) were constructed from the data in our initial HDF file with no further processing. In all other cases, the validation and test sets were held to the condition that they contain equal numbers of complexes from each percentile of the relevant binding affinity distribution to ensure a fair assessment of the model's performance across the full range of affinities. Other than this requirement and the condition that there is no overlap between any of the sets within a given protocol, the validation set used for testing on the PDBbind v.2016 core set as well as both the validation and test sets for the evaluation on lower-quality data were randomly generated from the PDBbind v.2020 general set, and the remaining complexes were used for training. For our 10-fold cross-validation, the sets were held to the above requirements in addition to membership in the PDBbind v.2020 refined set. Additionally, the ten test sets had no overlapping complexes. For each of the cross-validation splits, we utilized the Pybel software package[78] to compute SMILES strings from ligand MOL2 files and $T_c$ for each member of the test set with every other ligand in the PDBbind v.2020 refined set. Ligands not in the test set were removed if the $T_c$ between them and any ligand in the test set was greater than 0.7. Additionally, we asserted that the average $T_c$ between any ligand not in the test set and the collection of all test set complexes was less than 0.25. After this filtering process, the validation set was selected randomly from the remaining complexes while enforcing the equal distribution of binding affinities detailed above. All other refined-set complexes not filtered out by the similarity requirements were then used as training data.

## 7.4 Voxelization

Once training, validation, and test sets had been assembled in HDF format, they could immediately be used by the GCN components of the HAC-Net architecture. However, for use in the 3D-CNN, the atomic features must be voxelized into 4-dimensional grids with dimensions 48×48×48×19. We first aligned each protein-ligand complex to the center of a 3-dimensional voxel grid, then assigned each atomic feature vector to the voxel containing the center of the corresponding atom. On the rare occasion that two atom centers were positioned in the same voxel (0.22% of atoms), we summed the corresponding features within that voxel. If an atom's center was outside of the 48 Å voxel grid (0.01% of atoms), we omitted its features from the voxelized data. These 4-dimensional arrays were collected for all members of each dataset, and these collections were saved as HDF files.

## 7.5 Trainings

The 3D-CNN component of HAC-Net was constructed and trained using PyTorch.[57,58] The trainings were carried out for 100 epochs with a batch size of 50 complexes, utilizing the MSE loss function and the Root Mean Square Propagation (RMSProp) optimizer,[86] with an initial learning rate of 0.0007. If the number of complexes in the training set was not divisible by the batch size, the last batch contained fewer than 50 complexes. To reduce bias during the training, the order of the complexes was randomly shuffled for each epoch. After each epoch, the model was evaluated with the validation data, allowing us to assess the propensity of the training for overfitting, and a checkpoint containing model parameters was saved. In addition, we computed the average correlation between predicted and true values on validation data as $(\text{Spearman}\,\rho + \text{Pearson}\,\text{r})/2$ at the end of each epoch, and selected for feature extraction the training checkpoint that corresponded to the highest average correlation. We then used the extracted features as input to train a set of fully-connected layers identical to those used in the feature extraction protocol other than the single distinction of using batch normalization with momentum of 0.3 for estimating the moving mean and moving variance rather than 0.1, and followed the same procedure to select the checkpoint to be used in HAC-Net.

The GCN components of HAC-Net were constructed and trained primarily using PyTorch Geometric.[53,54] It is important to note that HAC-Net contains two GCNs which have identical architectures and training protocols. Each GCN was trained for 300 epochs with a batch size of 7 complexes, utilizing the MSE loss function and Adam optimizer,[87] with a constant learning rate of 0.001. The order of complexes was randomized for each epoch, and if the number of training complexes was not divisible by the batch size, the leftover complexes were simply discarded for that epoch. The GCNs were evaluated on the validation data after each training epoch, and the checkpoint that corresponded to the highest average correlation on the validation set was selected for each GCN.

## Data and Software Availability

All of our software, as well as all training, validation, and test sets used in this work are available as open source at https://github.com/gregory-kyro/HAC-Net/. Additionally, the HACNet Python package is published to PyPI at https://pypi.org/project/HACNet/.

## Author Contributions

GWK, RIB designed research; GWK, RIB developed software; GWK, RIB published the Python package; GWK, RIB performed research; GWK, RIB, VSB analyzed data; and GWK, RIB wrote the paper. All authors have given approval to the final version of the manuscript.

‡GWK and RIB contributed to this work equally

## Funding Sources

National Institutes of Health: Grants R01GM136815 (VSB) and 5T32GM00828335 (GWK)

## Acknowledgments

We acknowledge financial support from the National Institutes of Health under Grants R01GM136815 (VSB) and 5T32GM00828335 (GWK). VSB also acknowledges high-performance computer time from the National Energy Research Scientific Computing Center and from the Yale University Faculty of Arts and Sciences High Performance Computing Center.

## Abbreviations

HAC-Net, hybrid attention-based convolutional neural network; ML, machine learning; MLP, multi-layer perceptron; CNN, convolutional neural network; 3D, 3-dimensional; GCN, graph convolutional network; SE, squeeze-and-excitation; ReLU, rectified linear unit; GG-NN, gated graph neural network; GRU, gated recurrent unit; NMR, nuclear magnetic resonance; CASF-2016, 2016 Comparative Assessment of Scoring Functions; RMSE, root-mean-square error; MAE, mean absolute error; TM, template modeling; NW,

Needleman-Wunsch; $T_c$, Tanimoto coefficient; ECFP4, extended-connectivity fingerprint up to four bonds; MSE, mean squared error; ACC2, atomic charge calculator II; RMSProp, root mean square propagation.

# Supporting Information

## List of Figures and Tables

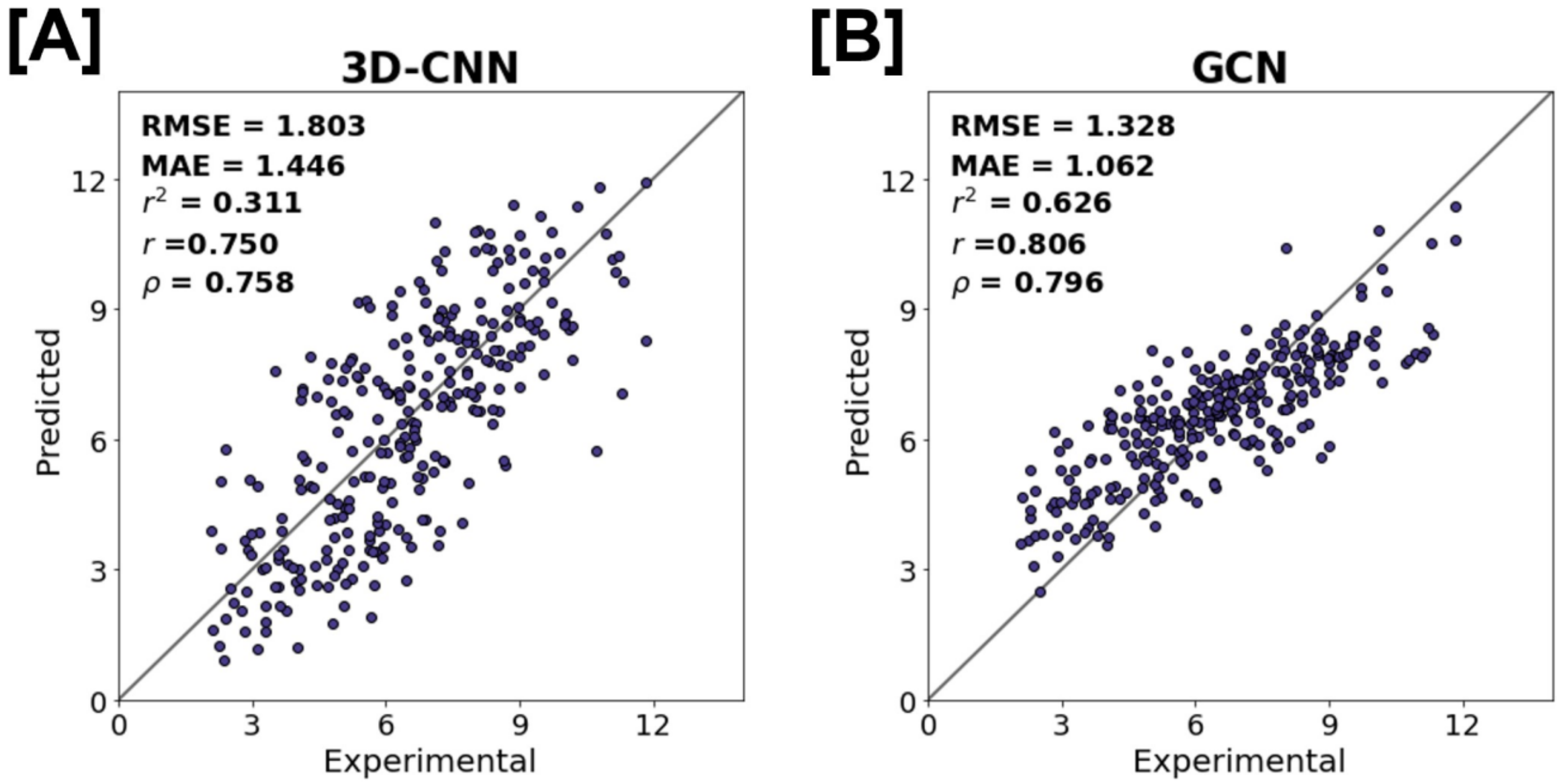

**Figure S1.** Correlation scatter plots depicting predictions of HAC-Net subcomponents on experimental p$K_D$ values of protein-ligand complexes in the PDBbind v.2016 core set. (A) 3D-CNN and (B) GCN are shown. Root-mean-square error (RMSE), mean absolute error (MAE), $r^2$, Pearson r, and Spearman $\rho$ are shown on plots.

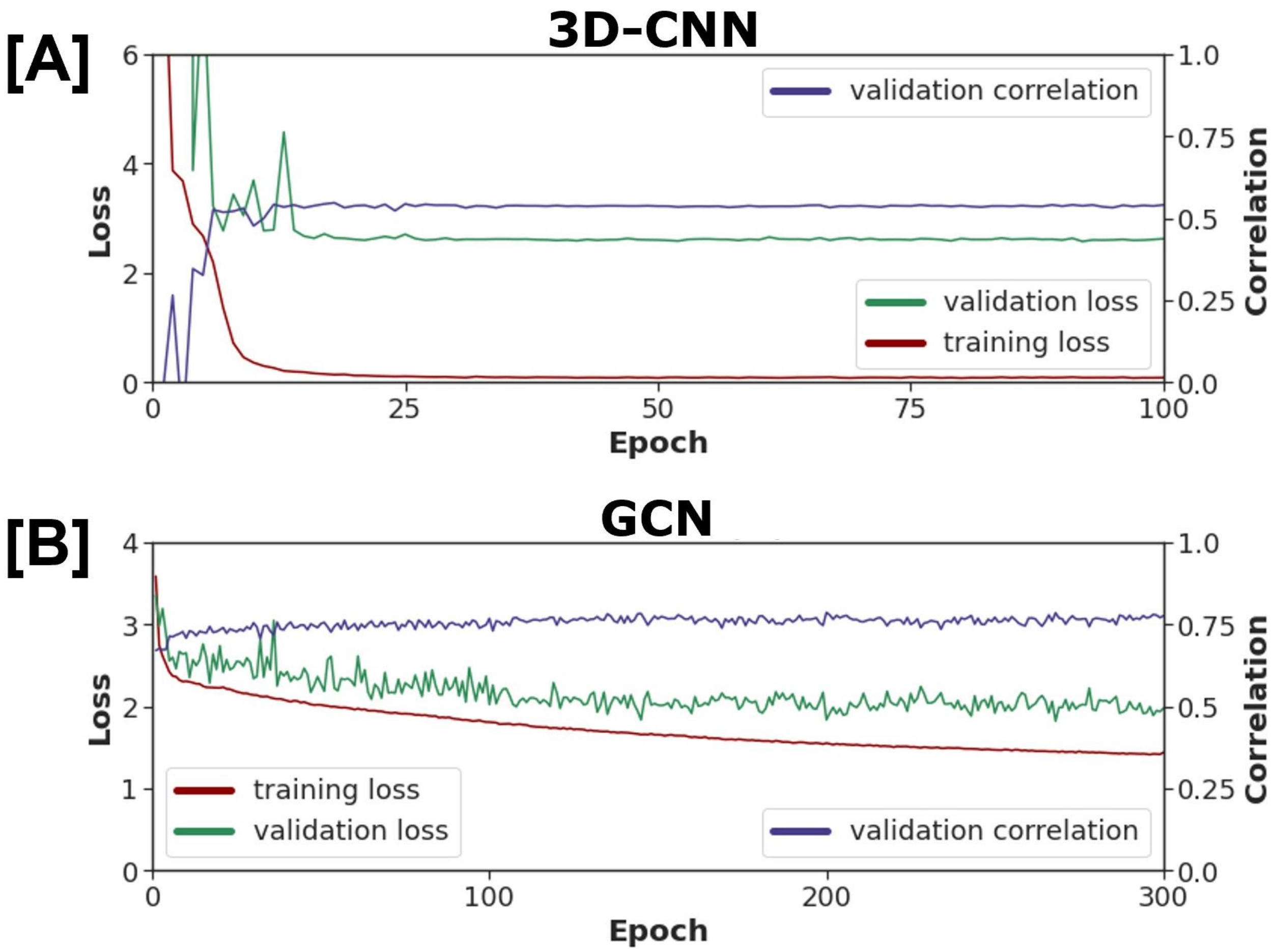

**Figure S2.** Representative learning curves for testing on the PDBbind v.2016 core set. Validation and training loss (left y-axis) and average correlation ((Spearman $\rho$ + Pearson r)/2) on the validation set (right y-axis) are shown as a function of epoch for the (A) 3D-CNN feature extraction and (B) one of the GCNs.

**Table S3.** Performance of HAC-Net on the Comparative Assessment of Scoring Functions (CASF)-2016 ranking, docking, and screening tests for protein ligand complexes in the CASF-2016 test set.

| | Ranking | | | Docking | | | Screening | | | | | |
|---|---|---|---|---|---|---|---|---|---|---|---|---|
| **Model** | Spearman $\rho$ | PI | Kendall $\tau$ | SR Top 1 | SR Top 2 | SR Top 3 | SR 1% F/R | SR 5% F/R | SR 10% F/R | Mean EF 1% | Mean EF 5% | Mean EF 10% |
| ***HAC-Net*** | 0.705 | 0.731 | 0.611 | 0.368 | 0.572 | 0.702 | 0.088/0.042 | 0.211/0.109 | 0.386/0.168 | 2.24 | 1.91 | 1.71 |

[a]We assess ranking power with mean spearman $\rho$, predictive index (PI) and Kendall $\tau$ across all 57 proteins, and docking power with success rate (SR), where a complex is marked as a success if the root-mean-square deviation (RMSD) of the top 1, 2 and 3 identified ligands is below a preset cutoff of 2.0 Å. To assess screening power, we calculate the SR of identifying the highest-affinity binder among the 1%, 5%, and 10% top-ranked ligands for each target protein in the test set (F: forward) and the SR of identifying the highest-affinity binder among the 1%, 5%, and 10% top-ranked proteins for each target ligand (R: reverse). Additionally, we utilize the mean enhancement factor (EF) among all proteins in the test set. This entire procedure is outlined by Su et al. (*J. Chem. Inf. Model.* 2019, 59, 2, 895–913)

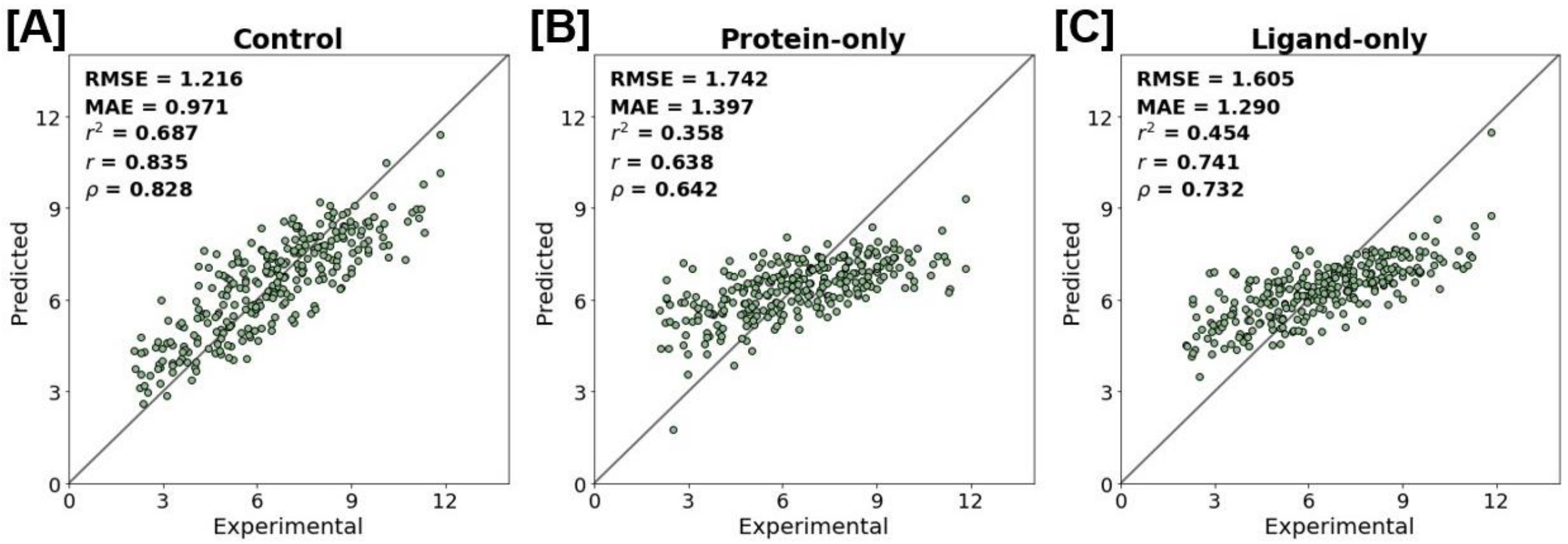

**Figure S4.** Correlation scatter plots depicting the performance of HAC-Net on the protein-ligand complexes of the PDBbind v.2016 core set compared to protein-only and ligand-only trainings and tests. Root-mean-square error (RMSE), mean absolute error (MAE), $r^2$, Pearson r, and Spearman $\rho$ are shown. Predictions of experimental p$K_D$ values are shown on the (A) protein-ligand complex data (control), (B) protein-only data, and (C) ligand-only data.

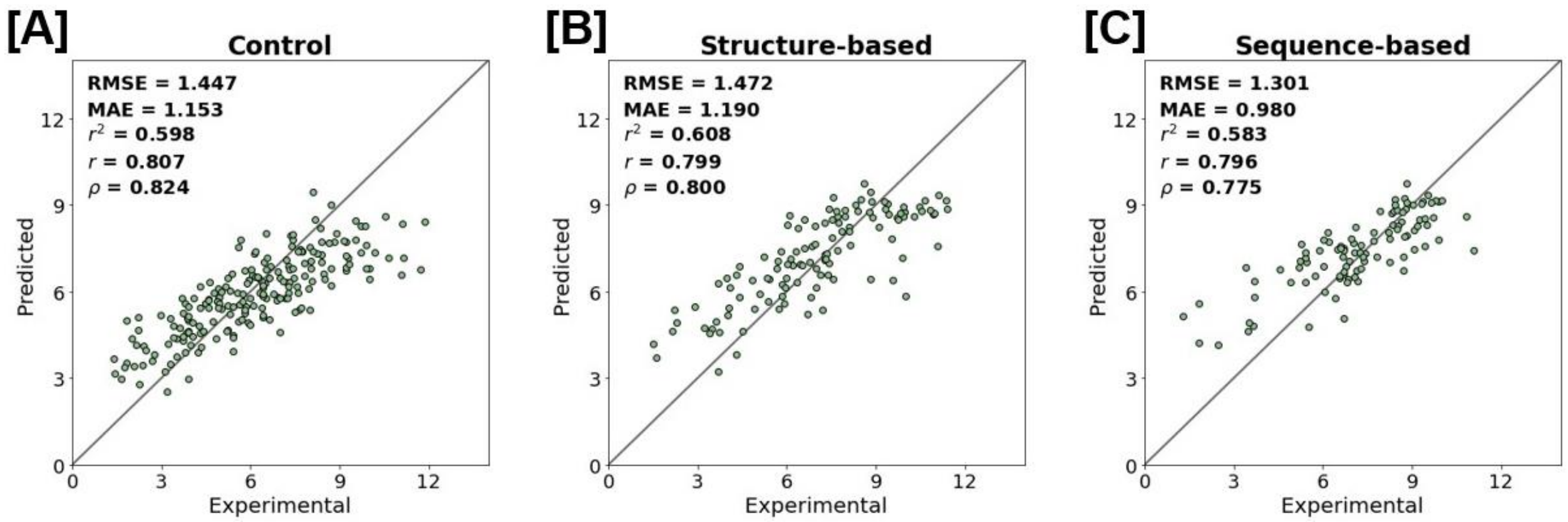

**Figure S5.** Correlation scatter plots depicting generalizability of HAC-Net across protein structure and sequence. Root-mean-square error (RMSE), mean absolute error (MAE), $r^2$, Pearson r, and Spearman $\rho$ are shown for predictions of experimental p$K_D$ values for complexes in the A) PDBbind v.2007 core set (Control), (B) test set based on protein structure-dissimilarity (Structure-based), and (C) test set based on protein sequence-dissimilarity (Sequence-based).

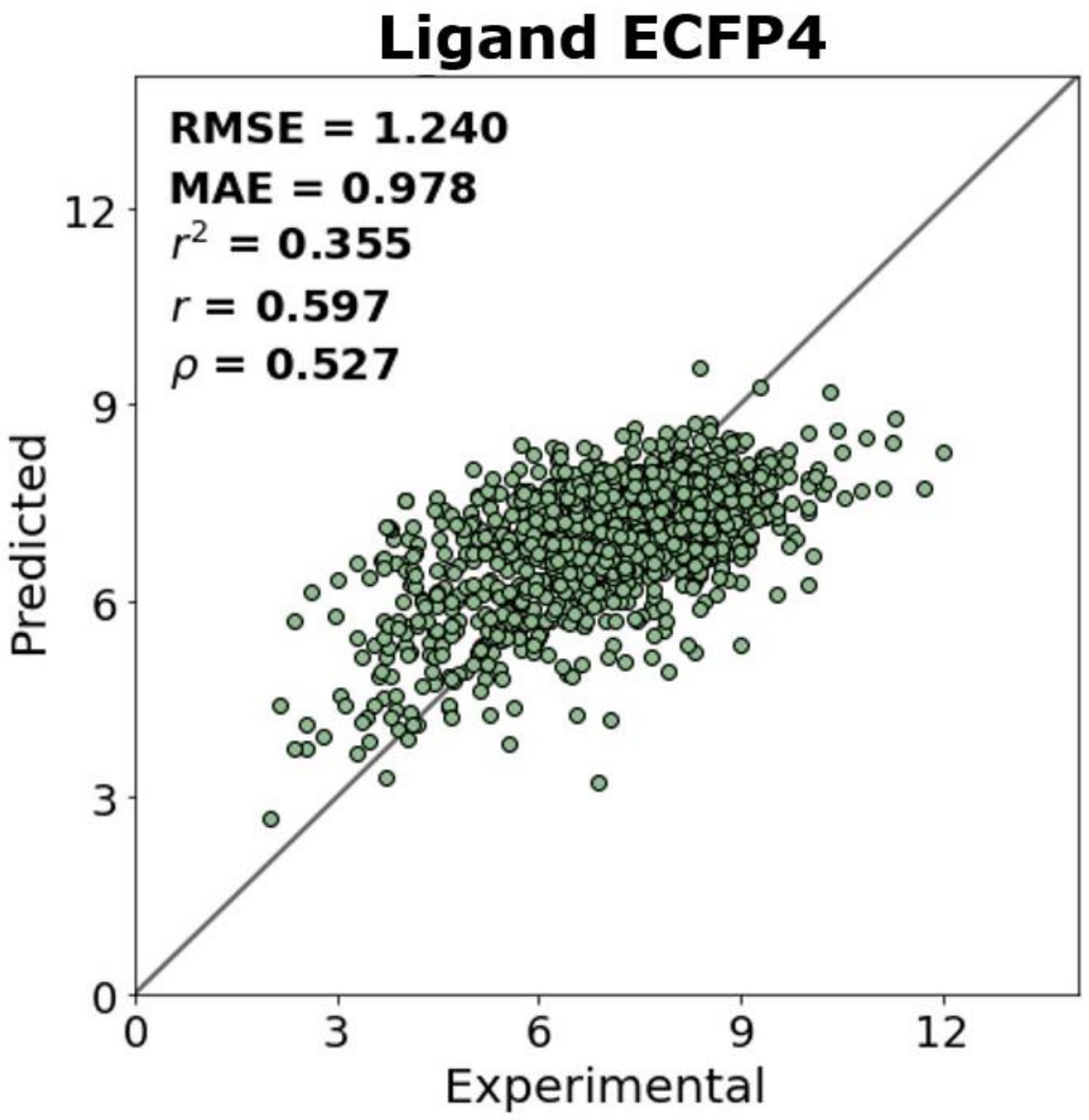

**Figure S6.** Correlation scatter plot depicting generalizability of HAC-Net based on ligand extended-connectivity fingerprints across four bonds (ECFP4s). Root-mean-square error (RMSE), mean absolute error (MAE), $r^2$, Pearson r, and Spearman $\rho$ are shown for predictions of experimental p$K_D$ values.

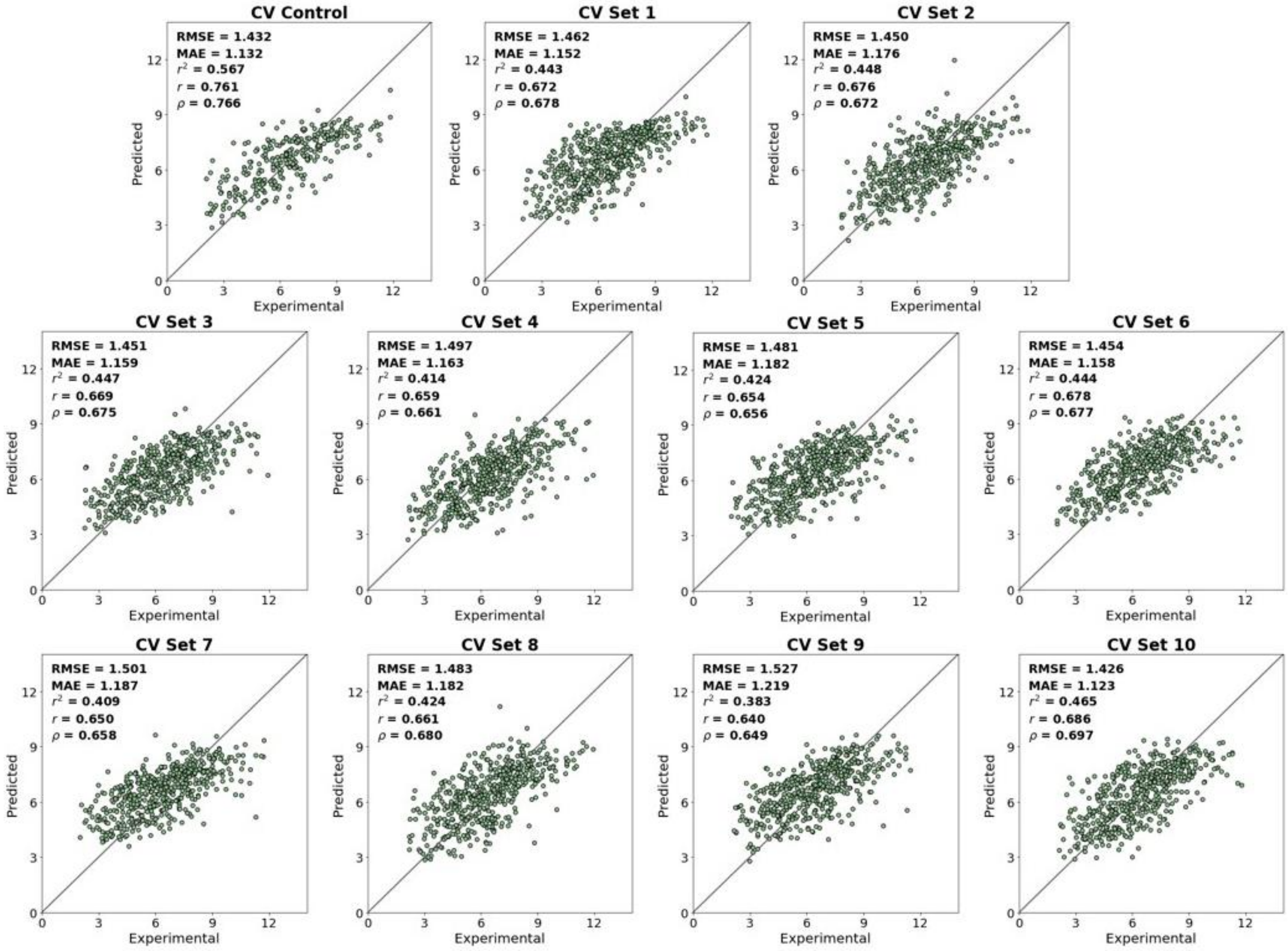

**Figure S7.** Correlation scatter plots depicting performance of HAC-Net on 10-fold cross-validation based on Tanimoto coefficient ($T_c$) cutoff applied to ligand SMILES strings. Root-mean-square error (RMSE), mean absolute error (MAE), $r^2$, Pearson r, and Spearman $\rho$ are shown for predictions of experimental p$K_D$ values on the PDBbind v.2016 core set (CV Control), as well as the ten cross-validation test sets.

# For Table of Contents Use Only

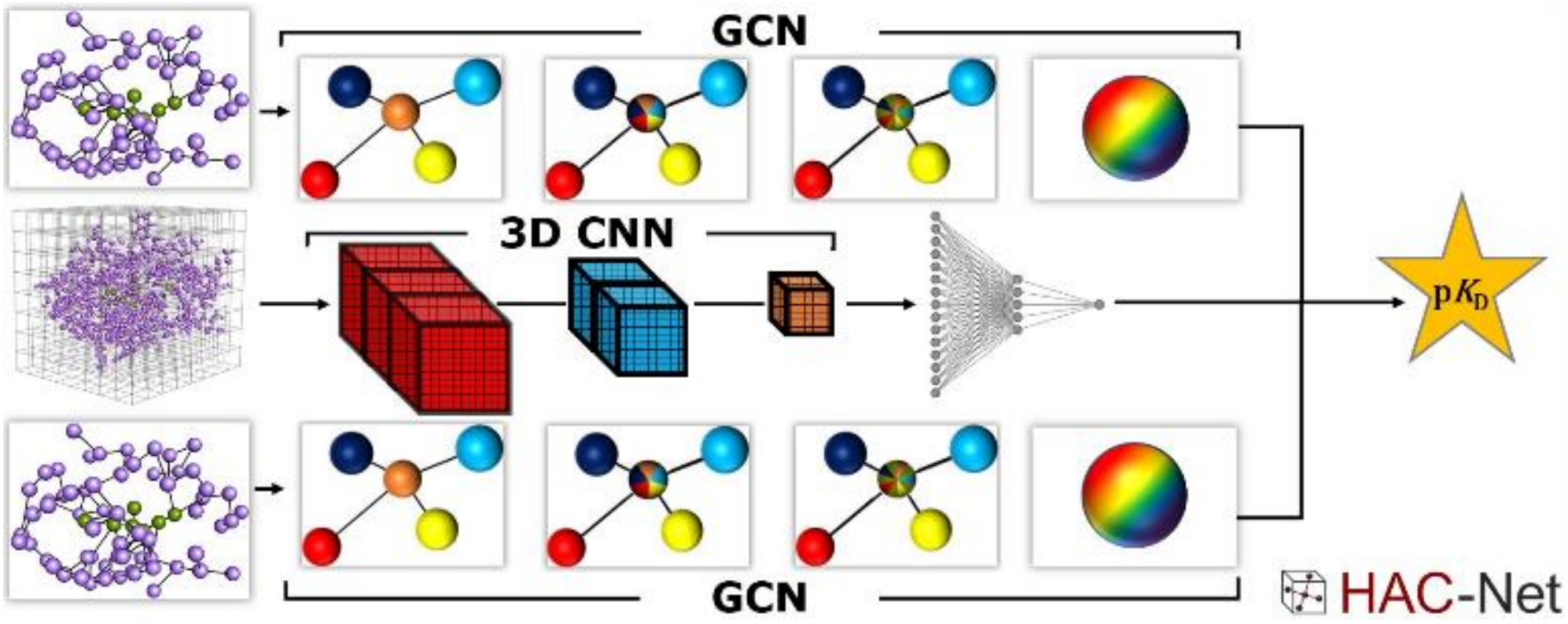